# Particle-Hole Ansatz in the Jaynes-Cummings-Hubbard Model

Moorad Alexanian

*Department of Physics and Physical Oceanography*
*University of North Carolina Wilmington, Wilmington, NC 28403-5606*

Email: alexanian@uncw.edu

**Abstract.** A recurrence relation ansatz between annihilation operators applied to the hopping interaction term of the Jaynes-Cummings-Hubbard model (JCHM) reduces the JCHM to that of the ordinary Jaynes-Cummings model (JCM), albeit, with a boson energy depending on the hopping strength. This allows us to calculate the phase diagram for the Mott-to-superfluid phase transition and the critical hopping strength as a function of the detuning.

## 1. Introduction

Quantum phase transitions have been observed for ultracold atom in a periodic lattice [1], for instance, where a Mott-insulator-superfluid phase transition has been observed in ultracold Rubidium atoms trapped in a three-dimensional optical lattice [2]. The dynamics of the transition of polaritons is determined by the repulsive interaction between the atoms on the same lattice site (localization) and the gain in kinetic energy when a polariton tunnels from site-to-site (delocalization). A prominent example is the superfluid-Mott insulator transition of polaritons in an array of coupled quantum electrodynamics (QED) cavities as described by the JCHM [3–5]. The JCHM is quite useful in the study of the interface of quantum optics and condensed matter physics.

In the present paper, we consider a one-dimensional JCHM and apply a recurrence relation ansatz to the hopping term that reduces the JCHM Hamiltonian to that of the JCM albeit with a boson energy depending on the hopping strength. This paper is structured as follows. In Sec. II, we introduce the recurrence relation ansatz for the hopping term in the JCHM and treat separately the particle and the hole ansatz. In Sec. III, we review the important results in the JCM and present the reduced JCHM for both the particle and the hole Hamiltonian. In Sec. IV, we present numerical results from our analytic expressions for the boundaries between Mott-insulating and superfluid phases and give an analytic expression of the critical value of the hopping strength as a function of the detuning. Finally, Sec. V summarizes our results.

## 2. Jaynes-Cummings-Hubbard model

The JCHM is a many-body quantum system modeling the quantum phase transition of light with the Hamiltonian given by ($\hbar = 1$)

$$H_{JCH} = \omega_c \sum_j a_j^\dagger a_j + \omega_z \sum_j \frac{\sigma_{jz}+1}{2} + g \sum_j (a_j^\dagger \sigma_{j-} + \sigma_{j+} a_j) - J \sum_j (a_j^\dagger a_{j+1} + a_{j+1}^\dagger a_j), \qquad (1)$$

where $\sigma_{j\pm}$ are Pauli operators for the two-level atom in the $j$th cavity and $\omega_c$ the frequency of the cavity modes. The coupling $J$ is the tunneling rate between nearest-neighbor cavities, $g$ the coupling strength between the qubit coupled to the cavity mode, $\omega_z$ the level splitting of the qubits, and $a_j$ ($a_j^\dagger$) the annihilation (creation) operator of the $j$th cavity mode, $j = 0, \pm 1, \pm 2, \ldots$ .

We consider two simple analytic approximations to the hopping interaction part of the Hamiltonian (1) that consist of the following two recurrence relations ansatz for the particle-hole excitations:

$$a_{j+1} = (\sqrt{3}+1)a_j - a_{j-1} \qquad \text{(particle)} \qquad (2)$$

$$a_{j+1} = (\sqrt{3}-1)a_j + a_{j-1} \qquad \text{(hole)} \qquad (3)$$

**A. Particle ansatz**

We apply the recurrence relation (2) and obtain, on summing over the infinite sites of the cavities, where $k = 0, \pm 1, \pm 2, \ldots$, the following recurrence relation for the sums

$$\sum_j (a_{j+k+1}^\dagger a_j + a_j^\dagger a_{j+k+1}) = (\sqrt{3}+1)\sum_j (a_{j+k}^\dagger a_j + a_j^\dagger a_{j+k}) - \sum_j (a_{j+k-1}^\dagger a_j + a_j^\dagger a_{j+k-1}), \qquad (4)$$

where we have used

$$\sum_j (a_{j-k}^\dagger a_j + a_j^\dagger a_{j-k}) = \sum_j (a_{j+k}^\dagger a_j + a_j^\dagger a_{j+k}). \qquad (5)$$

$$\sum_j (a_j^\dagger a_{j+1} + a_{j+1}^\dagger a_j) = (\sqrt{3}+1)\sum_j a_j^\dagger a_j \qquad (6)$$

The following results for the sums (4) for $k = 0, 1$ are, respectively
and

In fact, all hopping terms, in addition to the nearest neighbor hopping (6) and the second nearest neighbor hopping (7) are proportional to the total number of bosons.

$$\sum_j (a_{j+2}^\dagger a_j + a_j^\dagger a_{j+2}) = 2(\sqrt{3}+1)\sum_j a_j^\dagger a_j. \qquad (7)$$

**B. Hole ansatz**

We apply the recurrence relation (3) and obtain, on summing over the infinite sites of the cavities, the following recurrence relation for the sums

$$\sum_j (a_{j+k+1}^\dagger a_j + a_j^\dagger a_{j+k+1}) = -(\sqrt{3}-1)\sum_j (a_{j+k}^\dagger a_j + a_j^\dagger a_{j+k}) - \sum_j (a_{j+k-1}^\dagger a_j + a_j^\dagger a_{j+k-1}), \qquad (8)$$

where we have used (5). The following results for the sums (8) for $k = 0, 1$ are, respectively

$$\sum_j (a_j^\dagger a_{j+1} + a_{j+1}^\dagger a_j) = -(\sqrt{3}-1)\sum_j a_j^\dagger a_j \qquad (9)$$

and

$$\sum_j (a_{j+2}^\dagger a_j + a_j^\dagger a_{j+2}) = -2(\sqrt{3}-1)\sum_j a_j^\dagger a_j. \qquad (10)$$

In fact, all distance-dependent hopping terms, in addition, to the nearest neighbor hopping (9) and the second-nearest-neighbor hopping (10) terms, are all proportional to the total number of bosons.

The renormalization of the hopping strength with the addition of distance-dependent hopping terms can result in a vanishing hopping term in the resulting Hamiltonian. For instance, adding to the Hamiltonian (1) a second-nearest-neighbor interaction with hopping parameter $+J/2$ will give rise to no hopping term in the resulting Hamiltonian when using (9) and (10). Same results apply for the particle ansatz when using (6) and (7).

## 3. Model Hamiltonian

### A. Jaynes-Cummings model

The Hamiltonian of the JCM is [6])

$$H_{JC} = \omega_c \sum_j a_j^\dagger a_j + \omega_z \sum_j \frac{\sigma_{jz}+1}{2} + g \sum_j (a_j^\dagger \sigma_{j-} + \sigma_{j+} a_j). \tag{11}$$

The eigenstates of $H_{JC}$ includes the ground state with zero energy $|g_0\rangle = |0,\downarrow\rangle$ and the dressed states or polariton doublets, viz., $|n, \pm\rangle$, with $n = 1, 2, 3, \ldots$ excitations with the atom in the excited state $|n-1,\uparrow\rangle$ or with the atom in the ground state $|n,\downarrow\rangle$,

$$\begin{aligned} |n,+\rangle &= \cos(\theta/2)|n,\downarrow\rangle + \sin(\theta/2)|n-1,\uparrow\rangle \\ |n,-\rangle &= \sin(\theta/2)|n,\downarrow\rangle - \cos(\theta/2)|n-1,\uparrow\rangle \end{aligned} \tag{12}$$

with corresponding eigenvalues

$$E_{n,\pm} = (n-1/2)\omega_c + \omega_z/2 \pm \chi(n)/2, \tag{13}$$

where

$$\sin(\theta/2) = \sqrt{[1 - \Delta/\chi(n)]/2}, \tag{14}$$

$$\chi(n) = \sqrt{\Delta^2 + 4g^2 n} \tag{15}$$

and detuning

$$\Delta = \omega_c - \omega_z. \tag{16}$$

### B. Reduced Jaynes-Cummings-Hubbard model

Here we consider the JCHM Hamiltonian for the cases of particle and hole when the above ansatz is applied.

### C. Particle ansatz

The JCHM Hamiltonian (1) becomes, under the assumed ansatz (2) and with the aid of (6),

$$H_P = \left(\omega_c - (\sqrt{3}+1)J\right) \sum_j a_j^\dagger a_j + \omega_z \sum_j \frac{\sigma_{jz}+1}{2} + g \sum_j (a_j^\dagger \sigma_{j-} + \sigma_{j+} a_j), \tag{17}$$

which is the JCM, albeit, with a modified boson energy $\omega_c \to \omega_c - (\sqrt{3}+1)J$.

The eigenstates and eigenvalues of $H_p$ are given by those of the JCM, albeit with the replacement $\omega_c \to \omega_c - (\sqrt{3}+1)J$ and $\Delta \to \bar{\Delta} - (\sqrt{3}+1)J$ so. We then have

$$|n,+\rangle_P = \cos(\theta_P/2)|n,\downarrow\rangle + \sin(\theta_P/2)|n-1,\uparrow\rangle$$
$$|n,-\rangle_P = \sin(\theta_P/2)|n,\downarrow\rangle - \cos(\theta_P/2)|n-1,\uparrow\rangle \quad (18)$$

and

$$E^P_{n,\pm} = (n-1/2)(\omega_c - (\sqrt{3}+1)J) + \omega_z/2 \pm \chi_P(n)/2, \quad (19)$$

where

$$\sin(\theta_P/2) = \sqrt{\left(1 - (\Delta - (\sqrt{3}+1)J)/\chi_P(n)\right)/2}, \quad (20)$$

$$\chi_P(n) = \sqrt{\left(\Delta - (\sqrt{3}+1)J\right)^2 + 4g^2 n}, \quad (21)$$

and detuning

$$\Delta = \omega_c - \omega_z. \quad (22)$$

**D. Hole ansatz**

The JCHM Hamiltonian (1) becomes, under the assumed ansatz (3) and with the aid of (9),

$$H_H = \left(\omega_c + (\sqrt{3}-1)J\right)\sum_j a_j^\dagger a_j + \omega_z \sum_j \frac{\sigma_{jz}+1}{2} + g\sum_j (a_j^\dagger \sigma_{j-} + \sigma_{j+} a_j), \quad (23)$$

which is the JCM, albeit, with a modified boson energy $\omega_c \to \omega_c + (\sqrt{3}-1)J$.

The eigenstates and eigenvalues of $H_H$ are given by those of the JCM, albeit with the replacement $\omega_c \to \omega_c + (\sqrt{3}-1)J$ and so $\Delta \to \Delta + (\sqrt{3}-1)J$. We then have

$$|n,+\rangle_H = \cos(\theta_H/2)|n,\downarrow\rangle + \sin(\theta_H/2)|n-1,\uparrow\rangle$$
$$|n,-\rangle_H = \sin(\theta_H/2)|n,\downarrow\rangle - \cos(\theta_H/2)|n-1,\uparrow\rangle \quad (24)$$

and

$$E^H_{n,\pm} = (n-1/2)(\omega_c + (\sqrt{3}-1)J) + \omega_z/2 \pm \chi_H(n)/2, \quad (25)$$

where

$$\sin(\theta_H/2) = \sqrt{\left(1 - (\Delta + (\sqrt{3}-1)J)/\chi_H(n)\right)/2}, \quad (26)$$

$$\chi_H(n) = \sqrt{\left(\Delta + (\sqrt{3}-1)J\right)^2 + 4g^2 n}, \quad (27)$$

and

$$\Delta = \omega_c - \omega_z. \quad (28)$$

## 4. Mott insulating lobes

In order to determine the phase diagram of the JCHM at zero temperature, viz., the two Mott lobes, one must first determine the chemical potentials which requires knowledge of the energies of both particles and holes.

### A. Particle ansatz

We consider the case of $n = 1$ atoms inside each cavity and zero detuning, that is, $\Delta = 0$, and so the chemical potential is

$$(\mu^P - \omega_c)/g = (E^P_{2,-} - E^P_{1,-} - \omega_c)/g = -(\sqrt{3}-1)(J/g) - \sqrt{(2+\sqrt{3})(J^2/2g^2)+2} + \sqrt{(2+\sqrt{3})(J^2/2g^2)+1} \quad (29)$$

### B. Hole ansatz

$$(\mu^H - \omega_c)/g = (E^H_{1,-} - E^H_{0,-} - \omega_c)/g = (\sqrt{3}-1)(J/2g) - \sqrt{(2-\sqrt{3})(J^2/2g^2)+1}, \quad (30)$$

where $E^H_{0,-} = 0$.

In Fig. 1, the two equations (29) and (30) define the upper and lower phase boundary, respectively, in the quantum phase diagram of the superfluid-Mott insulator transition and is shown by the green dash-dot. The point where the two lines meet, viz., $\mu^P = \mu^H$, corresponds to $J_c/g = 0.193$ and represents an upper limit for the size of the first Mott lobe. In Fig. 1, we also show the density-matrix renormalization-group (DMRG) (red cross) data points [7], the mean field (box) [4, 8-10], and the fermion approximation [11] (blue dash).

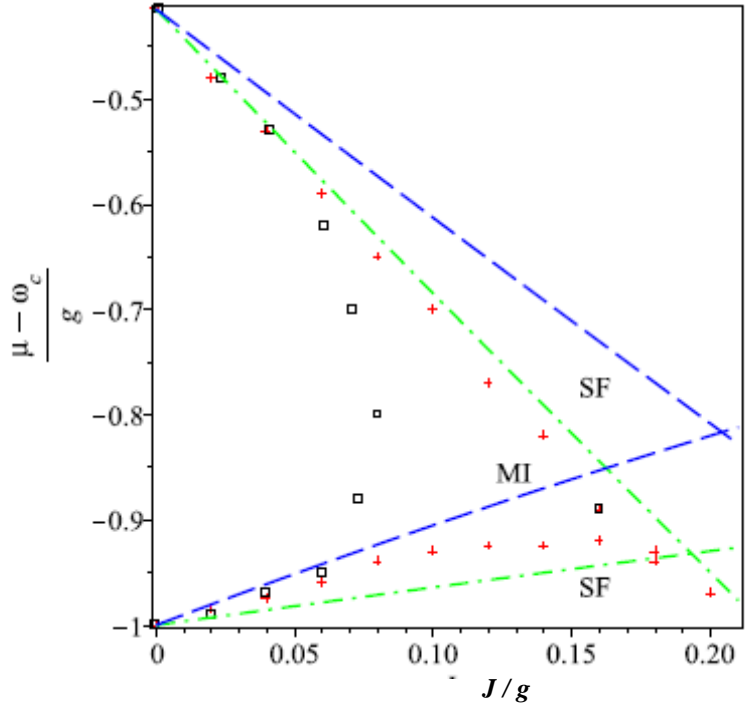

**Fig. 1.** Ground-state phase diagram, Mott insulator (MI) and superfluid (SF), for the one-dimensional JCHM determined from the data points [7] obtained from the density-matrix renormalization-group (DMRG) (red cross). Mean field [4, 8–10] (box). Fermion approximation [11] (blue dash). Present result (green dash-dot) with the critical point at $J_c/g = 0.193$.

Fig. 2 shows the critical hopping strength $J_c/g$ as a function of the detuning $\Delta/g$, which becomes vanishingly small for large negative detuning. In this dispersive regime, bosons and qubits barely interact with each other. Note also that in the limit of large detuning, i.e., $|\Delta| \to \infty$ the nature of the polariton quasiparticles becomes qubit-like for both particles and holes. The latter behavior follows also in the limit $|J| \to \infty$.

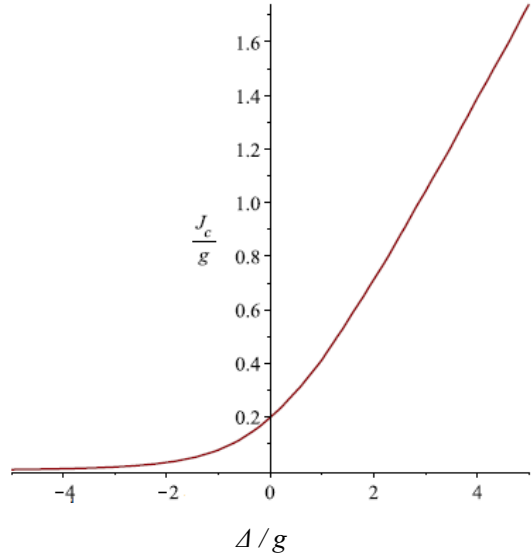

*Δ / g*

**Fig. 2.** Plot of the critical hopping strength $J_c/g$ versus the detuning $\Delta/g$, with $J_c/g = 0.193$ at $\Delta/g = 0$.

## 5. Conclusions

We have introduced a novel recurrence-relation ansatz for the annihilation (creation) operators and applied it to the hopping term of the JCHM that reduced the JCHM to that of the JCM albeit with a boson energy which is now dependent on the hooping amplitude. The application of the corresponding ansatz to the hole and particle boundaries of the quantum phase diagram separating the superfluid-Mott insulator transition region of polaritons is in good accord with the numerical results obtained with the aid of the density-matrix renormalization-group. It should be remarked that such numerical data points, as contrasted to analytical results, are obtained for finite systems whereas our results apply to a truly infinite array of cavities.